# Fuzzy Logic-based Implicit Authentication for Mobile Access Control

Feng Yao, Suleiman Y. Yerima, BooJoong Kang, Sakir Sezer
Centre for Secure Information Technologies,
Queen's University Belfast
Belfast, Northern Ireland, UK
fyao02@qub.ac.uk, s.yerima@qub.ac.uk, b.kang@qub.ac.uk, s.sezer@ecit.qub.ac.uk

*Abstract*—In order to address the increasing compromise of user privacy on mobile devices, a Fuzzy Logic based implicit authentication scheme is proposed in this paper. The proposed scheme computes an aggregate score based on selected features and a threshold in real-time based on current and historic data depicting user routine. The tuned fuzzy system is then applied to the aggregated score and the threshold to determine the trust level of the current user. The proposed fuzzy-integrated implicit authentication scheme is designed to: operate adaptively and completely in the background, require minimal training period, enable high system accuracy while provide timely detection of abnormal activity. In this paper, we explore Fuzzy Logic based authentication in depth. Gaussian and triangle-based membership functions are investigated and compared using real data over several weeks from different Android phone users. The presented results show that our proposed Fuzzy Logic approach is a highly effective, and viable scheme for lightweight real-time implicit authentication on mobile devices.

*Keywords—fuzzy logic, implicit authentication; mobile access control; behavior-based authentication*

## I. INTRODUCTION

Mobile devices have become increasingly ubiquitous and indispensable in modern society. They are widely adopted to enhance work efficiency, achieve convenient lifestyle and accomplish day-to-day tasks involving monetary transactions and storage or transfer of sensitive information. As a result, an increasing number of malicious activities are targeting mobile devices thus compromising the security and privacy of user data. Since mobile devices are particularly vulnerable to loss or physical theft, an efficient and intelligent access control strategy should be implemented to secure sensitive data at all times. Access control is usually enabled by authentication of the user (i.e. the system should verify that the user is legitimate).

Traditionally, users opt to use a PIN or password to provide access control for simplicity and ease of use. However, a weak password is subject to brute force or dictionary attack. Long and complicated passwords score low on usability, especially on mobile devices. A recent survey [1] revealed that users choose convenience over password-based security. To make matters worse, users prefer to keep themselves perpetually logged-in to sensitive online accounts and applications unless required by the application to log in every time, which makes user sensitive information vulnerable to theft (despite the use of passwords). Other authentication techniques such as requesting SMS authentication message and adding OTP (one time password) tokens could make the user data more secure, but these techniques are highly inconvenient and tokens are easily prone to loss. Biometric-based authentication is summarized in [2] which includes face and iris, fingerprint, voice/speech. It is reflective of who you are and more user appealing [3]. However, biometric-based authentication schemes can still be compromised in some circumstances and are computationally expensive for memory and power constrained devices.

As the traditional authentication methods do not suffice from the security and usability perspective, *implicit authentication* has been explored recently as an alternative to consistently protect the phone from unauthorized usage while provide maximum convenience. For example, [4-8] proposed schemes based on user behavior patterns or context. Although the results are promising, maintaining a balance between accuracy, adaptiveness, and practical feasibility is still an unresolved challenge. In our previous work [9], we proposed and investigated an event-driven scheme that incorporates user behavior awareness through data from everyday interaction with the phone. Motivated by the potential performance enhancing properties of Fuzzy Logic to our previous approach, in this paper we propose and evaluates a Fuzzy Logic based authentication scheme to provide more intelligent access control on mobile devices. Hence, in order to provide real-time, lightweight and transparent operation in practical real-life scenarios, we consider easily derived user behavior features for user profiling and Fuzzy Logic enhanced decision module to infer the user trust level. The trust level can further be used to activate or trigger explicit authentication (e.g. password, pin, voice, etc.) when the trust level indicates a considerable deviation from the usual or routine user behavior. Compared to previous works, key contributions in our proposed approach include:

1) Transparent, adaptive and event-driven authentication scheme with integration of Fuzzy Logic to enable implicit access control.

2) Extensive experiments on Fuzzy Logic based access control schemes with real behavioral data collected from Android devices.

In the rest of the paper, Section II reviews background of Fuzzy Logic while Section III describes our proposed authentication scheme. Section IV presents the evaluation of the scheme using real captured data from two Android devices. Section V summarizes the related work in mobile user behavior-based authentication. The paper is concluded in section VI.

## II. BACKGROUND

Fuzzy Logic [10] is a form of many-valued logic to handle the concept of partial truth. Fuzzy Logic is capable of reasoning and making rational decisions in an environment of imprecision, uncertainty and incompleteness of information. Because Fuzzy Logic allows for the inclusion of vague human assessments in computing problems, it has been widely adopted in the development of intelligent systems for decision making, identification, pattern recognition, optimization and control. Considering the effectiveness of Fuzzy Logic on decision making, Fuzzy Logic based decision making is incorporated to our behavior-based system to offer implicit authentication with more intelligence.

Fig. 1. Fuzzy-Logic implementation process

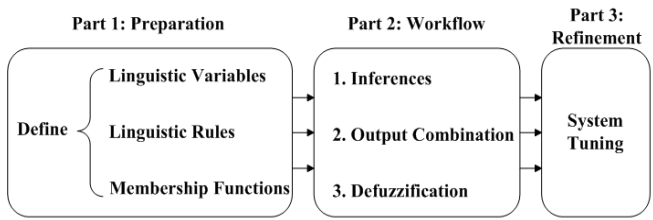

Briefly, an implementation of Fuzzy Logic usually comprises of three major stages as shown in Fig. 1. The first stage in implementing fuzzy system is to decide exactly what is to be controlled and how. Thus, linguistic variables and rules need to be defined to represent fuzzy system's operating parameters and regulate how these input parameters influence fuzzy output set. Appropriate membership function is selected as a graphical representation of the magnitude of participation of each input. Stage 2 depicts the basic workflow of Fuzzy Logic once the first stage is completed. In stage 3, the optimal performance is normally accomplished by tuning the fuzzy system, by techniques such as changing the rules, changing the shape of the input/output membership functions, or adding additional grades to the input/output membership functions.

Fuzzy Logic has been successfully applied in numerous fields for decades. And unsurprisingly, we can still find its adoption in lots of current scientific researches across multiple disciplines. A Fuzzy Logic based intelligent system for detecting and eliminating potential fires in the engine and battery compartments of a hybrid electric vehicle is proposed in [11]. Fuzzy Logic has also been used to recognize human action [12] and for dynamic tanker steering control [13]. The authors in [14] presents a Fuzzy Logic based resource-adaptive admission control for network management while [15] utilized Fuzzy Logic to optimize access network selection in wireless networks. Also, a Fuzzy Logic weight estimation method is proposed in [16] to improve security posture in a biometric-enabled co-authentication system in Android platform. Unlike the previous works, this paper proposes a novel application of Fuzzy Logic for behavior based authentication. As mentioned before, Fuzzy Logic is capable of making rational decisions with incomplete knowledge of current situation which makes it a perfect component to enhance mobile access control with intelligence.

## III. PROPOSED SCHEME

The Fuzzy Logic based authentication scheme proposed in this paper firstly models the user profile through selected user behavior-based features. Then a scoring algorithm is applied to compute the score of the current user while a threshold is adaptively computed along with the score. Finally, based on the knowledge of the score and threshold, the implemented fuzzy system is able to infer the trust level of the user, below which an explicit authentication process can be triggered or activated to provide access control. Our scheme is aimed at operation without dependence on a user training set as much as possible. Hence, from several possible threshold computation methods investigated in [9], EWMA (Exponentially weighted moving average) is selected as this threshold computation method to enable minimal training phase.

### A. Feature-based user behavior modelling

The proposed approach utilizes features derived from events depicting everyday user routines or patterns of behavior. It aims to use easily obtainable, readily available and commonly found data that can be extracted from most phones in order to construct a user's profile from routine phone usage. Hence, we consider the following as sources of data for our implicit authentication scheme:

1) Incoming SMS
2) Outgoing SMS
3) Incoming Call
4) Outgoing Call
5) Browser history
6) WIFI history

The experiments presented in this paper are based on these six features as sources of data. We plan to extend the scheme to include data from additional sources like application usage, phone state etc. in the future in order to further enrich the user profiling capability. However, our previous investigation has shown these six to be sufficient for high accuracy performance.

The workflow of proposed authentication scheme is presented in Fig. 2. Every time a feature-related event occurs, the event is captured by the event monitoring module. After that, the scoring module computes a score for the associated feature (e.g. if there is an incoming or outgoing SMS, the score for the incoming or outgoing SMS is calculated), and then a new aggregate score is computed. Afterward, the threshold computation module computes the threshold from the score(s). Because the computed score and threshold captures the users' latest behavior, the fuzzy system utilizes the distance between current score and previous score (*SSD*), and the distance between current score and threshold (*STD*) as the input variables.

The Fuzzy Logic based decision module is activated each time new values of *SSD* and *STD* are received. By processing the input variables within the pre-defined Fuzzy Logic system, the fuzzy system is able to generate a crisp output value **indicating the trust level of current user**. With the knowledge of the trust level and pre-defined threshold, the system is capable of determining whether the user passes the implicit authentication test or not. Failure of the implicit authentication test triggers an explicit authentication mechanism as explained earlier.

Fig. 2. System Block Diagram

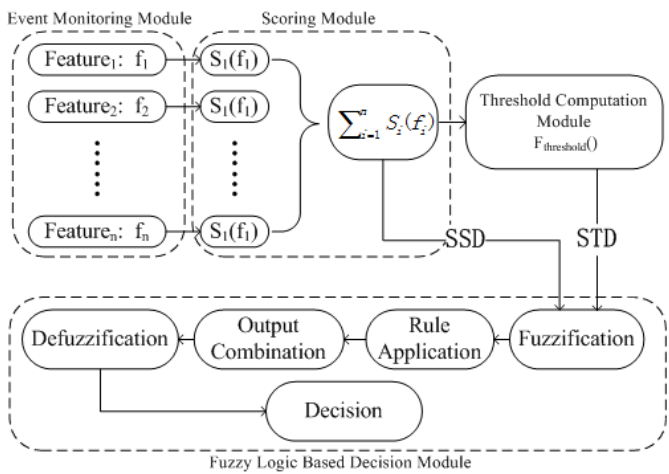

The features used to model user profile can be represented as:

$$Features : (f_1, f_2, f_3, ..., f_n) \quad (1)$$

The functions to compute the score for each feature are represented as:

$$Functions : (S_1, S_2, S_3, ..., S_n) \quad (2)$$

Thus, the aggregate score *AS* computed by this model can be represented as:

$$AS = \sum_{i=1}^{n} S_i(f_i) \quad (3)$$

Since the scheme is event driven, an aggregate score is calculated with each occurrence of a feature-related event. Afterward, the aggregate score is utilized to compute the threshold. Thus:

$$Threshold_i = F_{threshold}(AS_i, Threshold_{i-1}) \quad (4)$$

Where $F_{threshold}$ denotes the algorithm used to compute a suitable decision threshold. After the threshold is computed, the scoring module and threshold computation module send *SSD* and *STD* as inputs into Fuzzy Logic based decision module, where the trust level of current user can be inferred and final authentication decision can be made.

*B. Scoring algorithm*

A user profile comprises of a number of features which are close reflections of current user behavior. These features are employed to generate a final aggregate score to quantify the familiarity of current user activities. Table I illustrates how the score is allocated for each condition of different features.

TABLE I. SCORE ALLOCATION FOR EACH FEATURES

|  | Contact List | Top 5 | Duration |
|---|---|---|---|
| IncomingCall | 4 | 6 | 10 |
| OutgoingCall | 5 | 7 | 8 |
| IncomingSMS | 4 | 6 | -- |
| OutgoingSMS | 6 | 6 | -- |
| WIFIHistory | -- | 12 | -- |
| BrowserHistory | -- | 5 per domain | -- |

As it is shown in table I, the score for each feature is calculated based on different conditions. The score for each feature is certainly higher if more conditions are met. On discovering new feature-related user event, the system will check: (1) if the number is in the user's contact list (for incoming/outgoing SMS/Call), (2) if the number, or the WIFI host name, or the recently visited domains are in the top 5 list, (3) if the duration of incoming/outgoing call is longer than a certain period of time.

We maintain the top 5 list table dynamically based on 'the total number of occurrences' and 'the time elapse between the current computation and last computation' as discussed in detail in our previous paper [9]. Hence, the top 5 list is capable of updating itself adaptively in terms of the user behavior drift. A long duration call is considered a highly probable indication of normal usage and this is taken into consideration in the score computation for incoming/outgoing call. A score of 5 is assigned to each currently visited domain if they are found in top 5 list. If no browser activity is detected, the new score for browser history will be the previous score minus 5. The range for the score of browser history is limited from -5~20. A damping factor is introduced in the computation of the score for each feature in order to compensate for long periods of inactivity and allow for gradual decrease in the score.

*C. EWMA_SD_BLOCK threshold computation*

This threshold computation method is based on standard deviation (SD) and exponentially weighted moving average (EWMA). As shown in our previous work [9], EWMA_SD_BLOCK requires minimal training phase while providing high adaptivity compared to other threshold computation schemes. Firstly, it uses a consecutive output aggregate score as training dataset to compute the $threshold_1$. Secondly, it employs EWMA [17] method to compute a new threshold adaptively as shown below.

$$\begin{cases} Threshold_1 = mean - (standard\_deviation) \\ Threshold_t = \alpha \cdot ASBA_{t-1} + (1-\alpha) \cdot Threshold_{t-1}, \quad t > 1 \end{cases} \quad (5)$$

Here, *ASBA* stands for Aggregate Score Block Average. The system uses the average of the current block (block size *b* can be adjusted in the implementation) as input to *ASBA*. The new computed threshold is used as decision threshold for the next block. For example, assuming initial *AS* (aggregate score) index of current block is *k*, block size is set by *b*. The value of *ASBA* is expressed as follow:

$$ASBA = \frac{\left(\sum_{i=k}^{i=k+b} AS_i\right)}{b} \quad (6)$$

And the new computed threshold is used as detection threshold for next upcoming block from $AS_{k+b+1}$ to $AS_{k+2b+1}$. Because the optimal values of the operating parameters (block size *b*, coefficient *α* and training length *l*) was determined previously in [9], *b* = 6, *α* = 0.2, *l* = 5 days are employed throughout the experiments in this paper.

*D. Fuzzy Logic implementation-membership functions*

As shown in Fig. 2, Fuzzy Logic is integrated into the authentication system as a crucial component. This section presents the detailed implementation of fuzzy system's operating parameters (grades of input and output variables, and their membership functions).

Initially, the input membership functions for *SSD* and *STD* contain three grades which are 'negative', 'medium' and 'positive' as shown in Fig. 3. The output membership function contains three grades 'adversary', 'unknown' and 'legitimate' as well. Additionally, Input membership functions with five grades 'highly negative', 'negative', 'medium', 'positive', and 'highly positive' are also implemented as shown in Fig.4 (Gaussian) and Fig. 5 (Triangle) respectively. Finally, based on the implementation in Fig.4, Fig.6 adds two more grades into its output membership function along with the 'tweaked' input membership function to enhance the performance. Fig.7 is built upon Fig.5 in a similar way.

Fig.3. Gaussian membership functions for SSD (3), STD (3) and output (3)

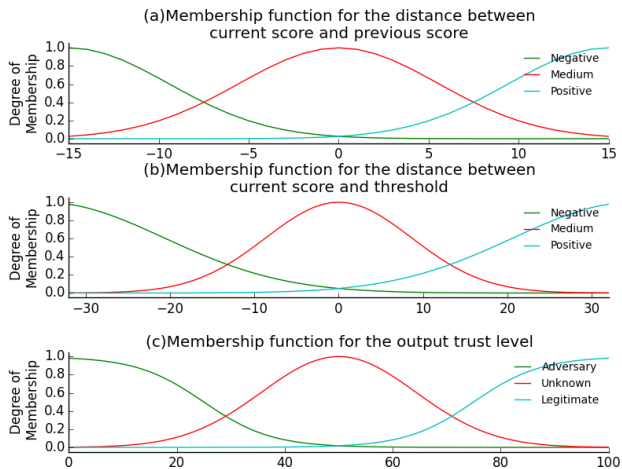

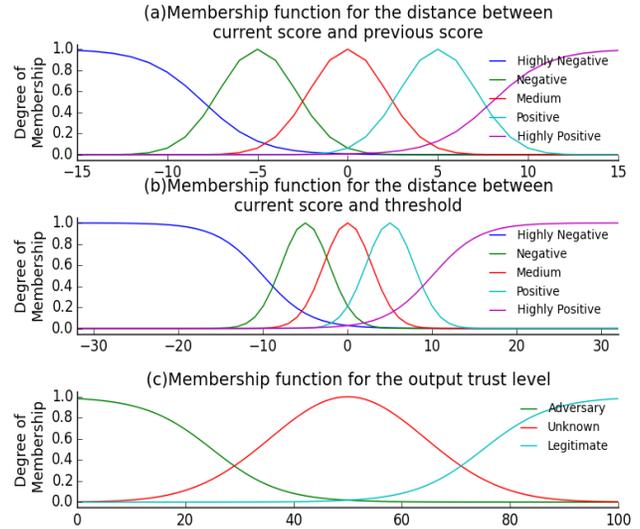

Fig.4. Gaussian membership functions for SSD (5), STD (5) and output (3)

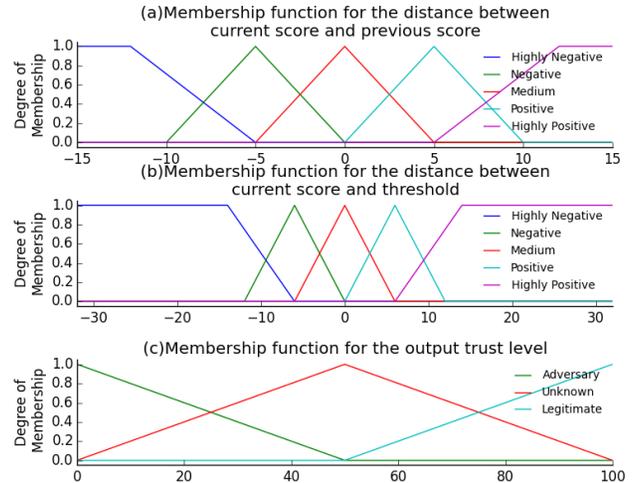

Fig.5. Triangle membership functions for SSD (5), STD (5) and output (3)

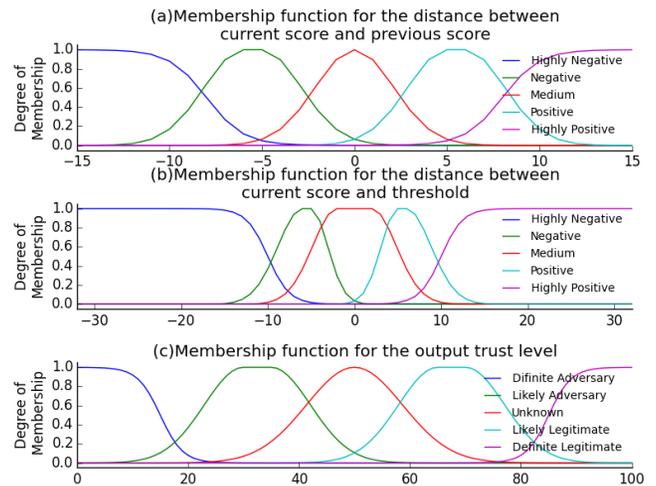

Fig.6. Gaussian membership functions for SSD (5), STD (5) and output (5)

Fig.7. Gaussian membership functions for SSD (5), STD (5) and output (5)

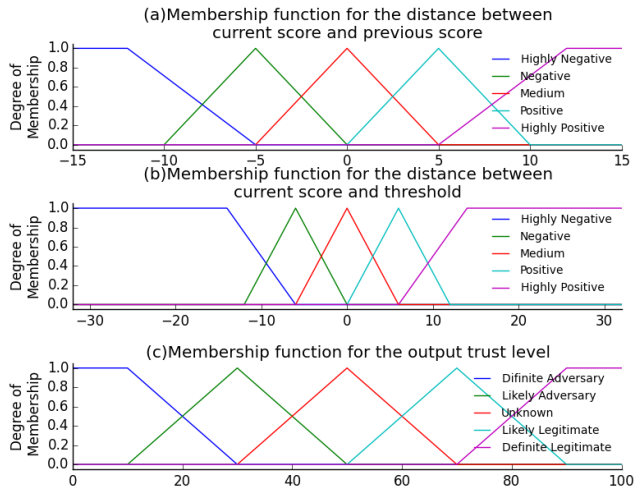

### E. Fuzzy Logic implementation-linguistic rules

The Fuzzy Logic based decision module applies a list of fuzzy IF-THEN rules as shown in Table II, III and IV. Table II is the rule matrix for three grades input and output membership functions (Fig.3). Table III is the rule matrix for five grades of input membership functions and three grades of output membership function (Fig.4 and 5). Table IV is the rule matrix for five grades input and output membership functions (Fig. 6 and 7). These rules map the input space to the output space in a linguistic manner. With the knowledge of the firing strength of each rule, the logic result for each rule should be combined in a proper way. ROOT-SUM_SQUARE is selected in our scheme since it is capable of giving the best weighted influence to all firing rules (as shown below).

$$Grade\_strength = \sqrt{\sum_i Rule_i^2} \quad (7)$$

With the combination of the results of the previous process, the defuzzification of the data into a crisp output value is accomplished by computing the 'fuzzy centroid' of the area.

$$Centroid = \frac{\int_s x \cdot y(x) dx}{\int_s y(x) dx} \quad (8)$$

Where S denotes the support of y(x) and y(x) is the result after applying firing strength of each grade into output membership function.

Table II. RULE MATRIX FOR FIG.3

| SSD \ STD | Negative | Medium | Positive |
|---|---|---|---|
| Negative | Adversary | Unknown | Legitimate |
| Medium | Unknown | Unknown | Legitimate |
| Positive | Unknown | Legitimate | Legitimate |

Table III. RULE MATRIX FOR FIG.4 and 5

| SSD \ STD | HN | N | M | P | HP |
|---|---|---|---|---|---|
| HN | A | A | U | U | U |
| N | A | U | U | L | L |
| M | A | U | L | L | L |
| P | U | U | L | L | L |
| HP | U | U | L | L | L |

(A: Adversary, U: Unknown, L: Legitimate)

Table IV. RULE MATRIX FOR FIG.6 and 7

| SSD \ STD | HN | N | M | P | HP |
|---|---|---|---|---|---|
| HN | DA | DA | LA | U | U |
| N | DA | LA | U | LL | LL |
| M | LA | U | LL | LL | LL |
| P | U | U | LL | DL | DL |
| HP | U | LL | LL | DL | DL |

(HN: Highly Negative, N: Negative, M: Medium, P: Positive, HP: Highly Positive
DA: Definite Adversary, LA: Likely Adversary, U: Unknown, LL: Likely Legitimate, DL: Definite Legitimate)

### IV. EXPERIEMNTS AND RESULTS

This section presents the experimental result and evaluation of proposed Fuzzy Logic based implicit authentication scheme. In particular, the impact different membership functions with different grades and shape have on the user recognition rate and the performance of adversary scenarios (i.e. simulated device theft) will be analyzed. An Android application is developed to extract the data based on the features described previously. In order to make the evaluation easier, the scoring module, threshold computation module and fuzzy module are implemented using Python scripts to emulate the logic of the Fuzzy Logic based authentication scheme. With the adoption of Fuzzy Logic, the user recognition rate is **dramatically increased from 55% to over 90%** compared with non-fuzzy authentication scheme.

In order to verify the performance of the proposed Fuzzy Logic based authentication scheme, two user data sets including respective adversary scenarios are collected over several weeks. Behavioral data regarding user A is obtained from Samsung Galaxy A5 device with Android version 4.4.4. Behavioral data regarding user B is obtained from Samsung Galaxy S3 with Android version 4.2. The developed application requires a minimal Android SDK version of 8.

In order to verify the efficiency and resilience of the proposed authentication scheme on dealing with normal and complex adversary scenarios, two adversary cases were investigated in the case of user A and one adversary case was investigated in the case of user B to emulate device theft. An example is given in Fig. 8 (Fig. 6 is used as membership functions in this example). The 'trigger threshold for trust level' is set as 30 in this example. The constant change of score and threshold reflects the attacker's behavior. Thus, with new input of *SSD* and *STD*, the fuzzy output can be computed to compare against 'trigger threshold for trust level'. When the 'fuzzy output trust level' is lower than the 'trigger threshold for trust level', it is considered as adversary and explicit authentication is prompted. In the case of user A, the attack starts at 15:00:00 (the device is taken by the attacker). The performance of legitimate use measurement accuracy is evaluated by observing user recognition rate. The performance of adversary detection is evaluated by observing two metrics: *NOC* (number of computations before the first detection as illegitimate use) and *elapsed time* in minutes (before first detection).

Fig.8. Example of adversary cases for user A and B

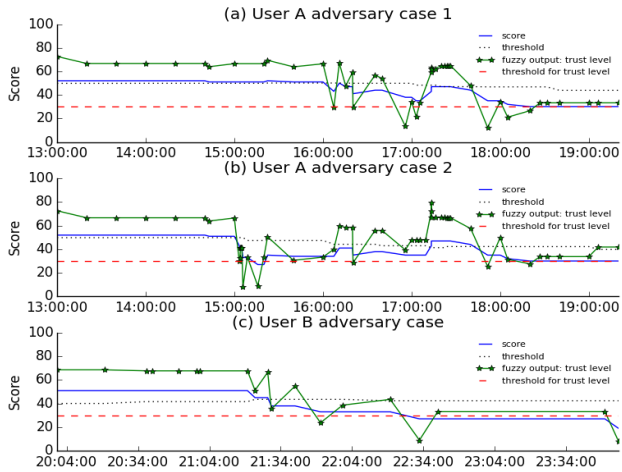

*User A Case 1:* The attacker takes the device and interacts with it continuously but not intensively (shown in Fig. 8 (a)). E.g. connecting to WIFI previously unknown to the device, browsing unfamiliar websites, calling/texting numbers unfamiliar to the device.

*User A Case 2:* This case represents a more intensive and immediate interaction with the device after it is taken from the legitimate user (shown in Fig. 8 (b)).

*User B:* The attack starts at 21:15:00. The device is stolen by the attacker on the subway (shown in Fig. 8 (c)).

Fig. 9 to 11 depict the results of the experiments undertaken to evaluate the accuracy and adversary detection performance (User Recognition Rate, NOC, Elapse time) of the developed Fuzzy Logic based authentication scheme.

Fig.9. Experimental results of data of user A with adversary case 1

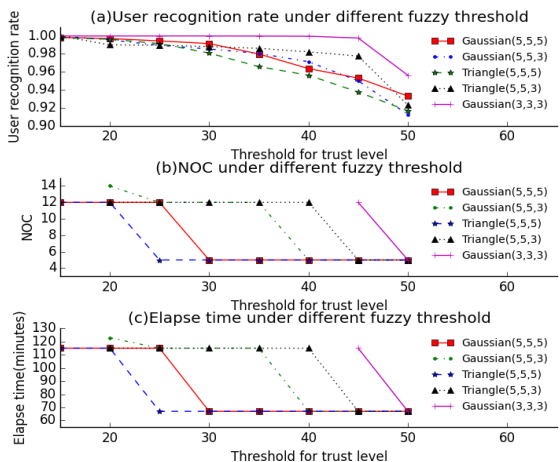

Fig.10. Experimental results of data of user A with adversary case 2

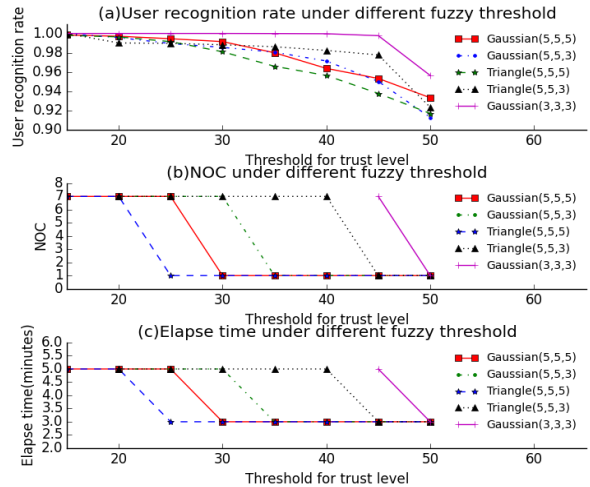

Fig.11. Experimental results of data of user B

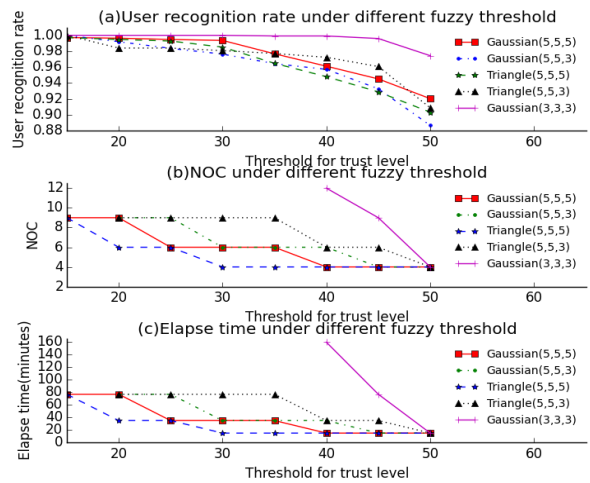

Gaussian(5,5,5) refers to Gaussian-based membership functions with 5 grades of SSD, STD and output trust level (as shown in Fig. 6). Triangle(5,5,3) refers to Triangle-based membership functions with 5 grades of SSD, STD and 3 grades of output trust level (as shown in Fig. 5). The rest are defined in a similar way

Fig. 9-11 are the results of the authentication process on the user A data set with adversary case 1, user A data set with adversary case 2 and user B data set respectively. The user recognition rate ranges from 90% to 100% for Gaussian and Triangle-based schemes in both user case A and B, which is acceptable from the perspective of convenience and usability. An interesting observation in three figures is that Gaussian(3,3,3) has the highest user recognition rate around 99%, but it has the worst adversary detection performance and is unable to detect the adversary case when the 'trigger threshold for trust level' is lower than certain value (45 for user A and 40 for user B). Hence, membership functions with additional grades are considered as alternatives to enhance the resilience of the proposed authentication scheme.

Fig. 9-11 illustrates that the optimal 'trigger threshold for trust level' is around 30-35 where the optimum balance between high user recognition rate and faster adversary detection are achieved. Thus, we use 35 as benchmark for 'trigger threshold for trust level' in the rest of evaluation.

Table V. PERFORMANCE COMPARISON IN USER A WITH ADVERSARY CASE 1

|  | Gaussian (5,5,3) | Triangle (5,5,3) | Gaussian (5,5,5) | Triangle (5,5,5) |
|---|---|---|---|---|
| URR | 98.05% | 98.62% | 97.99% | 96.58% |
| NOC | 12 | 12 | 5 | 5 |
| Elapse Time | 115 min | 115 min | 67 min | 67 min |

(URR: User recognition rate, trigger threshold for trust level = 35)

Table VI. PERFORMANCE COMPARISON IN THE CASE OF USER B

|  | Gaussian (5,5,3) | Triangle (5,5,3) | Gaussian (5,5,5) | Triangle (5,5,5) |
|---|---|---|---|---|
| URR | 96.44% | 97.72% | 97.68% | 96.48% |
| NOC | 6 | 9 | 6 | 4 |
| Elapse Time | 35 min | 77 min | 35 min | 15 min |

Through the observation of Table V and VI, it can be concluded that (1) Triangle-based and Gaussian-based membership function present a close performance in the proposed Fuzzy Logic based authentication scheme. (2) Generally, membership functions with more grades performs better than membership functions with less grades, especially on the performance of adversary detection.

## V. REALTED WORK

Related work on protecting data on mobile devices based on behavior or context can be found in the current literature. For example, a progressive authentication scheme is proposed in [5] which adopts face, voice recognition and other features to determine the level of confidence in a user's authenticity. Authors in [4] proposed a probability density function based implicit authentication by modelling and analyzing user behavioral patterns. This implicit authentication scheme requires considerable amount of training dataset and adopts a static threshold which limits its ability to detect drifts in user behavior. Another behavior based authentication scheme is proposed in [8] which utilize probability density function to model user behavior temporally and spatially. This scheme is able to automatically switch from training to deployment and activate retraining model on detecting behavior drift. However, this scheme requires frequent access to power-consuming sensor data (For example, GPS, mobile network location). A context-aware scalable authentication scheme is proposed in [6] which intends to minimize active authentication with the inference of surrounding context. Another context profiling framework is proposed in [7] which authenticates based on context variables like GPS readings, WIFI access point and Bluetooth devices. This framework is able to estimate the familiarity and safety of a context based on the context variables and use it to dynamically configure security policies. This scheme also allows user to provide feedback to calibrate the perceived safety of context. Unlike all of these previous works, our proposed scheme in this paper is based on a novel application of fuzzy logic for enhanced implicit authentication, with the advantages of operational transparency, adaptivity to behavior drift, and lightweight due to dependence on easily extractable event-based features.

## VI. CONCLUSION

This paper proposed and evaluated a Fuzzy Logic based implicit authentication scheme for mobile access control. The scheme is capable of maintaining a high recognition of legitimate user activities while enable timely detection of adversary user behavior by feeding dynamically computed aggregate score and threshold based parameters into Fuzzy Logic system to determine the trust status of current user activity. The proposed Fuzzy Logic based authentication scheme also operates transparently and adaptively, requires no user interaction, and triggers explicit authentication to prevent unauthorized access in adversarial scenarios. Extensive experiments were conducted to investigate the efficiency of the proposed authentication scheme using real data collected from two users (during normal routine behavior) between February-April and July-August 2015 respectively. Afterwards, three adversary use cases were emulated to measure the effectiveness of illegitimate use detection.

Through the analysis of experimental results, it can be concluded that it is feasible and practical to fulfill an intelligent access control strategy from readily available and easily obtainable everyday phone data. Besides, this work is further extensible based on Fuzzy Logic-centric structure. For example, location or application usage based metrics can be easily integrated with another specially designed membership function. Although our current results illustrate the viability of a high performance system based on six features, future work could explore the use of additional features to extend the range of user behavior data, other possible behavior profiling algorithms or intelligent decision support systems could also be explored for similar implicit mobile access control.

## REFERENCES

[1] Confident Technologies, "Survey Shows Smartphone Users Choose Convenience Over Security", http://confidenttechnologies.com/news_event/survey-shows-smartphone-users-choose-convenience-security/

[2] T. Stockinger, "Implicit authentication on mobile devices." In The Media Informatics Advanced Seminar on Ubiquitous Computing. 2011.

[3] M. El-Abed, R. Giot, B. Hemery, C. Rosenberger, "A study of users' acceptance and satisfaction of biometric systems." In Security Technology (ICCST), 2010 IEEE International CamaHan Conference on, IEEE, pp. 170-178, Oct. 2010.

[4] E. Shi, N. Yuan, M. Jakobsson, and R. Chow, "Implicit authentication through learning user behavior." In Information Security, Springer Berlin Heidelberg, pp. 99-113, 2011.

[5] O. Riva, C. Qin, K. Strauss, and D. Lymberopoulos, "Progressive Authentication: Deciding When to Authenticate on Mobile Phones." In USENIX Security Symposium, pp. 301-316, 2012.

[6] E. Hayashi, S. Das, S. Amini, J. Hong, and I. Oakley, "Casa: context-aware scalable authentication." In Proceedings of the Ninth Symposium on Usable Privacy and Security, ACM, pp. 3-13, 2013.

[7] A. Gupta, M. Miettinen, N. Asokan, and M. Nagy, "Intuitive security policy configuration in mobile devices using context profiling." In Privacy, Security, Risk and Trust (PASSAT), 2012 International Conference on and 2012 International Conference on Social Computing (SocialCom), IEEE, pp. 471-480, 2012.

[8] H. G. Kayacik, M. Just, L. Baillie, D. Aspinall, and N. Micallef, "Data driven authentication: On the effectiveness of user behaviour modelling with mobile device sensors." In Proceedings of the Third Workshop on Mobile Security Technologies (MoST), 2014.

[9] F. Yao, S. Y. Yerima, B. Kang, and S. Sezer, "Event-driven implicit authentication for mobile access control." 9th International Conference on Next Generation Mobile Applications, Services and Technologies (NGMAST), IEEE, 8 pages, 2014.

[10] L.A. Zadeh, "Is there a need for fuzzy logic?" Information Sciences 178, no. 13, pp. 2751-2779, Jul. 2008.

[11] M. S. Dattathreya, H. Singh, and T. Meitzler. "Detection and elimination of a potential fire in engine and battery compartments of hybrid electric vehicles." Advances in Fuzzy Systems, Vol. 2012, Article ID 687652, 11 pages, 2012.

[12] X. Q. Cao and Z. Q. Liu, "Type-2 Fuzzy Topic Models for Human Action Recognition." in Fuzzy Systems, IEEE Transactions on, vol.23, no.5, pp.1581-1593, Oct. 2015.

[13] N. Wang; M. J. Er; M. Han, "Dynamic Tanker Steering Control Using Generalized Ellipsoidal-Basis-Function-Based Fuzzy Neural Networks," in Fuzzy Systems, IEEE Transactions on, vol.23, no.5, pp.1414-1427, Oct. 2015.

[14] S. Y. Yerima, G. Parr, S. McClean and P. Morrow, "Adaptive Measure-Based Policy-Driven QoS Management with Fuzzy-Rule-based Resources Allocation." Future Internet, Vol. 4, pp. 646-671, Jul 2012.

[15] F. Kaleem, A. Mehbodniya, K. K. Yen and F. Adachi, "A Fuzzy Preprocessing Module for Optimizing the Access Network Selection in Wireless Networks." Advances in Fuzzy Systems, vol. 2013, Article ID 232198, 9 pages, 2013.

[16] V. N. Nguyen, V. Q. Nguyen, M. N. B. Nguyen and T. K. Dang, "Fuzzy Logic Weight Estimation in Biometric-Enabled Co-authentication Systems." Information and Communication Technology. Springer Berlin Heidelberg, pp. 365-374, 2014.

[17] S.Y. Yerima, "Implementation and Evaluation of Measurement-Based Admission Control Schemes Within a Converged Networks QoS Management Framework." International Journal of Computer Networks and Communications, IJCNC, Vol. 3, No. 4, July 2011. ISSN : 0974 - 9322[Online]; 0975 - 2293 [Print]